\documentstyle[12pt,psfig]{article}
\newcommand{\be}{\begin{eqnarray}}
\newcommand{\ee}{\end{eqnarray}}
\newcommand{\ba}{\begin{array}}
\newcommand{\ea}{\end{array}}
\newcommand{\half}{{\textstyle{\frac{1}{2}}}}
\newcommand{\partialslash}{\partial\hspace{-.5em}/\hspace{.15em}}
\newcommand{\Pslash}{P\hspace{-.5em}/\hspace{.15em}}

\newcommand{\kslash}{k\hspace{-.5em}/\hspace{.15em}}
\newcommand{\bfk}{{\bf k}}
\newcommand{\kint}{\int_{\Lambda_3}\!\frac{d^4 k}{(2\pi)^4}}

\begin{document}

\begin{center}

{\Large\bf
Decays of
excited strange mesons
in the extended NJL model.} \\[1.5cm]

{\large\bf M.K.\ Volkov
\footnote{
E-mail: volkov@thsun1.jinr.ru
}, V.L. Yudichev
\footnote{
E-mail: yudichev@thsun1.jinr.ru
}
}
 \\[0.4cm]
{\em Bogoliubov Laboratory of Theoretical Physics \\
Joint Institute for Nuclear Research \\
141980, Dubna, Moscow region, Russia}
\\[0.7cm]
\vskip0.5truecm

{\large\bf D.Ebert}
\footnote{
E-mail: debert@qft2.physik.hu-berlin.de
}
 \\[0.4cm]
{\em Institut f\"ur Physik \\
Humboldt-Universit\"at zu Berlin\\
Invalidenstrasse 110, D-10115
Berlin, Germany}
\end{center}
\vspace{1cm}

\begin{abstract}
\noindent
A chiral $U(3)\times U(3)$ Lagrangian, containing
besides the usual meson fields their first radial excitations,
is considered. The Lagrangian is derived by bosonization of the
Nambu--Jona-Lasinio (NJL) quark model with separable non-local
interactions.
The spontaneous breaking
of chiral symmetry is governed by the NJL gap equation. The first
radial excitations of the kaon,  $K^*$ and $\varphi$ are described
with the help of two form factors.
The values for the decay widths of the processes
$K^{*'} \to \rho K, K^{*'} \to K^* \pi $, $K^{*'} \to K \pi$,
$\varphi' \to K^* K$, $ \varphi' \to \bar KK $,
$K' \to K \rho$, $K' \to K^* \pi$ and $K'\to K 2\pi $
are obtained
in qualitative agreement with the experimental data.

\end{abstract}
\vfill
\newpage

\section{Introduction}
In our previous papers \cite{volk_96, volk_97, VEN} the chiral
quark model of the Nambu--Jona-Lasinio (NJL)  type with
separable non-local interactions has been proposed.
This model is a nonlocal extension of the standard
NJL model \cite{volkov_83,volk_86,ebert_86,voglweise,klev}.
The first radial excitations of the  scalar, pseudoscalar,
vector and axial-vector mesons were described with the
help of form factors corresponding to 3-dimensional
ground and excited state wave functions.
The meson masses, weak decay constants and a set of
 decay widths of nonstrange mesons were calculated.

The theoretical foundations for the choice of polynomial pion-quark form
factors were discussed in \cite{volk_96}
and it was shown that we can choose these form
factors in such a way that the mass gap equation conserves its usual form
and gives a solution with a constant constituent quark mass. Moreover, the
quark condensate does not change after including the excited states in the
model, because the tadpoles connected with the excited scalar fields vanish.
Thus, in this approach it is possible to describe radially excited mesons
above the usual NJL vacuum, preserving the usual mechanism of
chiral symmetry breaking. Finally, it has been shown that one can derive an
effective meson Lagrangian for the ground and excited meson states directly
in terms of local fields and their derivatives. A nonlocal separable
interaction is defined in the Minkowski space in a 3-dimensional (yet
covariant) way whereby form factors depend only on the part of the quark-
antiquark relative momentum transverse to the meson momentum.
This ensures  absence of spurious relative-time excitations
\cite{feynman_71}.

In the paper \cite{volk_97}, the meson mass spectrum for the ground and excited
pions, kaons and
the vector meson nonet in the $U(3) \times U(3)$ model of this
type has been obtained. By fitting the meson mass spectrum,
all parameters in this model are fixed. This then allows one to describe all the strong,
electromagnetic and weak interactions of these mesons without introducing
any new additional parameter.

In the paper~\cite{VEN}, it was shown that
this model satisfactorily describes two types of
decays. This concerns the
strong decays like $\rho \to 2 \pi, \pi' \to \rho \pi$, $\rho' \to 2\pi$
associated with divergent quark diagrams as well as the decays
$\rho' \to \omega \pi$ and $\omega' \to \rho \pi$ defined by anomalous quark
diagrams.
Here we continue the similar calculations for the description
of the decay widths of strange pseudoscalar and vector mesons.

The paper is organized as follows.
In section 2, we introduce the effective quark interaction
in the separable approximation and describe its bosonization.
In section 3, we derive the effective Lagrangian for the pions and kaons
and perform the diagonalization leading to the physical pion and kaon
ground and excited states. In section 4, we carry out the diagonalization
for the $K^*$- and $\varphi$-mesons.
In section 5, we give the parameters of our model
and  the masses of the ground and excited states of kaons,
$K^*$- and $\varphi$-mesons and the
weak decay constants $F_{\pi}$, $F_{\pi'}$,
$F_{K}$ and $F_{K'}$.
In section 6, we evaluate the decay widths of the processes
$K^{*'} \to K^* \pi$, $K^{*'}\to \rho K, K^{*'} \to K\pi,
\varphi' \to K^* K$ and
 $\varphi' \to \bar K K$.
In section 7, we calculate the decay widths of the processes $K'\to \rho K$,
$K'\to K^*\pi$ and $K'\to K\, 2\pi$.
The obtained results are discussed in section 8.

\section{$U(3)\times U(3)$ chiral Lagrangian with the excited meson
states }

We shall use a separable interaction, which is still
of current--current form, but allows for non-local
vertices (form factors) in the definition of the quark currents,
\be
L [\bar q, q] =
\int d^4 x \, \bar q (x) \left( i \partialslash - m^0 \right)
q (x) \; + \; \tilde L_{\rm int} ,
\label{L_NJL}
\ee
\be
\tilde{L}_{\rm int} &=&
\int d^4 x \sum_{a=0}^8\sum_{i = 1}^N \left[ \frac{G_1}{2}
\left[ j_{S,i}^a (x) j_{S,i}^a (x) +
j_{P,i}^a (x) j_{P,i}^a (x) \right]  \right. \nonumber \\
&&\left. - \frac{G_2}{2}  \left[ j_{V,i}^{a,\mu} (x) j_{V,i}^{a,\mu} (x) +
j_{A,i}^{a,\mu} (x) j_{A,i}^{a,\mu} (x) \right]  \right] ,
\label{int_sep}
\ee
\be
j^a_{S, i} (x) &=& \int d^4 x_1 \int d^4 x_2 \;
\bar q (x_1 ) F^a_{S, i} (x; x_1, x_2 ) q (x_2 ),
\label{j_S} \\
j^a_{P, i} (x) &=& \int d^4 x_1 \int d^4 x_2 \;
\bar q (x_1 ) F^a_{P, i} (x; x_1, x_2 ) q (x_2 ),
\label{j_P} \\
j^{a,\mu}_{V, i} (x) &=& \int d^4 x_1 \int d^4 x_2 \;
\bar q (x_1 ) F^{a,\mu}_{V, i} (x; x_1, x_2 ) q (x_2 ),
\label{j_V} \\
j^{a,\mu}_{A, i} (x) &=& \int d^4 x_1 \int d^4 x_2 \;
\bar q (x_1 ) F^{a,\mu}_{A, i} (x; x_1, x_2 ) q (x_2 ).
\label{j_A}
\ee
Here $m^0$ is the current quark mass matrix (we suppose that
$m_u^0 \approx m_d^0$);
$j^a_{S,P,V,A} (x)$ denote, respectively, the scalar,
pseudoscalar, vector and axial--vector currents of the
quark field;
$F^{a,\mu}_{U, i}(x; x_1, x_2 )$, \,
$i = 1, \ldots N$, are a set of non-local scalar,
pseudoscalar, vec\-tor and axial--vector quark ver\-tices
(in general momentum-- and spin--dependent),
which will be specified below.

Upon bosonization
we obtain~\cite{volk_96,volk_97}
\be
&&L_{\rm bos}(\bar q, q; \sigma, \phi, P, A) = \int d^4 x_1
\int d^4 x_2~ \bar q (x_1 ) \left[ \left( i \partialslash_{x_2}
- m^0 \right) \delta (x_1 - x_2 ) \right.      \nonumber \\
&&\qquad+ \int d^4 x \sum_{a=0}^8 \sum_{i = 1}^N
\left( \sigma^a_i (x) F^a_{\sigma , i} (x; x_1, x_2 ) +
\phi_i^a (x) F_{\phi , i}^a (x; x_1, x_2)  \right. \nonumber \\
&&\qquad\left.\left.
+ V_i^{a,\mu} (x) F_{V , i}^{a,\mu} (x; x_1, x_2) +
A_i^{a,\mu} (x) F_{A , i}^{a,\mu} (x; x_1, x_2) \right) \right] q (x_2 )
\nonumber \\
&&\qquad - \int d^4 x \sum_{i = 1}^N
\left[ \frac{1}{2G_1} \left( \sigma_i^{a\, 2} (x) +
\phi_i^{a\, 2} (x) \right)
- \frac{1}{2G_2} \left( V_i^{a,\mu\, 2} (x) + A_i^{a,\mu\, 2} (x)
\right) \right].
\label{L_sep}
\ee
This Lagrangian describes a system of local meson fields,
$\sigma_i^a (x)$, $\phi_i^a (x)$, $V^{a,\mu}_i (x)$, $A^{a,\mu}_i (x)$,
$i = 1, \ldots N$, which interact with
the quarks through non-local vertices. These fields
are not yet to be associated with physical particles,
which will be obtained after determining the vacuum and
diagonalizing the effective meson Lagrangian.
\par
In order to describe the first radial excitations of mesons (N = 2),
we take the form factors in the form (see \cite{volk_96} )
\be
F^a_{\sigma , 2} ({\bf k}) &=& \lambda^a f_a^P ({\bf k}),
\;\;\;\;\;
F^a_{\phi , 2} ({\bf k}) = i \gamma_5 \lambda^a f_a^P ({\bf k}),
\nonumber \\
F^{a,\mu}_{V , 2} ({\bf k}) &=& \gamma^\mu \lambda^a f_a^V ({\bf k}),
\;\;\;\;\;
F^{a,\mu}_{A , 2} ({\bf k}) = \gamma_5 \gamma^\mu \lambda^a
f_a^V ({\bf k}),
\label{ffs}
\ee
\be
f_a^U ({\bf k}) = c_a^U ( 1 + d_a {\bf k}^2 ).
\label{ff}
\ee
Where
$\lambda^a$ are the Gell--Mann matrices.
We consider here the form factors in the momentum space and in
the rest frame of the mesons (${\bf P}_{meson}$ = 0; $k$ and
$P$ are the relative and total momentum of the quark-antiquark
pair.). For the ground states of the mesons we choose the functions
$f_a^{U,0} ({\bf k})$ = 1.
\par
After integrating over the quark fields in
eq.(\ref{L_sep}), one obtains the effective Lagrangian of the
$\sigma_1^a , \sigma_2^a , \phi_1^a,  \phi_2^a, V_1^{a,\mu},
V_2^{a,\mu}, A_1^{a,\mu}$ and $ A_2^{a,\mu}$ fields.
\be
&&L(\sigma', \phi, V, A, \bar\sigma, \bar\phi, \bar V, \bar A) =\;\;\;\;\;\;\;~~~~~~~~~~
\nonumber \\
&&\qquad - \frac{1}{2 G_1} (\sigma_a^{'2} + \phi_a^2 + \bar\sigma_a^2 +
\bar\phi_a^2 )
+ \frac{1}{2 G_2} (V_a^2 + A_a^2 + \bar V_a^2 + \bar A_a^2 )
\nonumber \\
&&\qquad- i N_c \; {\rm Tr}\, \log [ i \partialslash - m^0 + (\sigma'_a
+ i \gamma_5  \phi_a +\gamma_\mu V^\mu_a +
\gamma_5 \gamma_\mu A^\mu_a   \nonumber \\
&&\qquad+ (\bar\sigma_a + i \gamma_5 \bar\phi_a) f_a^P + (\gamma_\mu \bar V^\mu_a +
\gamma_5 \gamma_\mu \bar A^\mu_a ) f_a^V ) \lambda^a ]
\label{12}
\ee
where we have put $\sigma_1=\sigma'$,
$\sigma_2=\bar\sigma$, $\pi_1=\pi$, $\pi_2=\bar\pi$ etc.
\par
 Let us define the vacuum expectation of the $\sigma'_a$ fields
\footnote{We can derive this form of the gap equation only
if the condition $\langle\bar\sigma_a\rangle_0=0$ is fulfilled
(see refs.~\cite{volk_96,volk_97,VEN} and eqs. (\ref{I_1^f}) )}
\be
<\frac{\delta L}{\delta\sigma'_a}>_0 &=& - i N_c \; {\rm tr} \kint
\frac{1}{( \rlap/k - m^0 + <\sigma'_a>_0 )}
- \frac{<\sigma'_a>_0}{G_1} \; = \; 0 .
\label{gap_1}
\ee
Introduce the new sigma fields whose vacuum expectations are
equal to zero
\be
\sigma_a = \sigma'_a - <\sigma'_a>_0
\label{sigma}
\ee
and redefine the quark masses
\be
m_a = m_a^0 - <\sigma'_a>_0.
\label{m^0}
\ee
Then eq. (\ref{gap_1}) can be rewritten in the form of the usual
gap equation
\be
m_i = m_i^0 + 8 G_1 m_i I_1 (m_i),  \;\;\;\;\;\; (i = u, d, s)
\label{gap}
\ee
where
\be
I_n (m_i) = -i N_c \; \kint \frac{1}{(m_i^2 - k^2)^n}
\label{I_n}
\ee
and $m_i$ are the constituent quark masses.
\par

\section{The effective Lagrangian for the ground and excited states
of the pions and kaons}
To describe the first excited states of all the meson nonets, it
is necessary to use three different slope parameters $d_a$ in the
form factors $f_a^U ({\bf k})$ (see eq. (\ref{ff}))
\be
f_{uu}^{P,V} ({\bf k}) = c_{uu}^{P,V} ( 1 + d_{uu} {\bf k}^2 ), \nonumber\\
f_{us}^{P,V} ({\bf k}) = c_{us}^{P,V} ( 1 + d_{us} {\bf k}^2 ), \nonumber\\
f_{ss}^{P,V} ({\bf k}) = c_{ss}^{P,V} ( 1 + d_{ss} {\bf k}^2 ).
\label{ffq}
\ee
Following our works \cite{volk_96,volk_97} we can fix the parameters $d_{uu},
d_{us} $ and $d_{ss}$ by using the conditions
\be
I_1^{f_{uu}^P} (m_u) = 0,\;\;~~~
I_1^{f_{us}^P} (m_u) + I_1^{f_{us}^P} (m_s) = 0, \;\;~~~
I_1^{f_{ss}^P} (m_s) = 0,
\label{I_1^f}
\ee
where
\be
I_1^{f_a..f_a} (m_i) = -i N_c \;
\kint \frac{f_a..f_a}{(m_i^2 - k^2)}.
\label{I_1^ff}
\ee
Eqs. (\ref{I_1^f}) allow us to conserve the gap
equations in the form usual for the NJL model (see eqs.
(\ref{gap})) because the tadpoles
with the excited scalar external fields do not contribute to
the quark condensates and to the constituent quark masses.
\par
Using eqs. (\ref{I_1^f}) we obtain close values for all $d_a$
\be
d_{uu} = - 1.784~{\rm GeV}^{-2},~~ d_{us} = - 1.7565~{\rm GeV}^{-2},~~
d_{ss} = - 1.727~{\rm GeV}^{-2}.
\label{d_a}
\ee
\par
Now let us consider the free part of the Lagrangian (\ref{12}).
For the pions and kaons we obtain
\be
L^{(2)} (\phi) &=&
\half \sum_{i, j = 1}^{2} \sum_{a = 1}^{7}
\phi_i^a (P) K_{ij}^{ab} (P) \phi_j^b (P) .
\label{L_2}
\ee
Here
\be
\sum_{a = 1}^{3} (\phi_i^a)^2 = (\pi_i^0)^2 + 2 \pi^+_i \pi^-_i,~~~
(\phi_i^4)^2 + (\phi_i^5)^2 = 2 K_i^+ K_i^-,\quad
(\phi_i^6)^2 + (\phi_i^7)^2 = 2 K_i^0 \bar K_i^0.
\label{phi^a}
\ee
The quadratic form $K_{ij}^{ab} (P)$, eq.(\ref{L_2}), is
obtained as
\be
&K_{ij}^{ab} (P) \equiv \delta^{ab} K_{ij}^a (P),&
\nonumber \\
&\displaystyle K_{ij}^a (P) = -~\delta_{ij} \frac{1}{G_1}
-~i~ N_{\rm c} \; {\rm tr}\, \displaystyle\kint \left[
\frac{1}{\displaystyle \kslash + \half\Pslash - m_q^a}
i\gamma_5 f_i^a
\frac{1}{\displaystyle \kslash - \half\Pslash - m_{q'}^a} i \gamma_5 f_j^a
\right],& \nonumber \\
&f_1^a \equiv 1, \hspace{2em} f_2^a \;\; \equiv \;\; f_a^P ({\bf k}).&
\label{K_full}
\ee
\be
m_q^a = m_u~~(a = 1,...,7);\quad
m_{q'}^a = m_u~~(a = 1,...,3);~~ m_{q'}^a = m_s~~ (a = 4,...,7).
\label{m_q^a}
\ee
$m_u$ and $m_s$ are the constituent quark masses ($m_u \approx m_d$).
The integral (\ref{K_full}) is evaluated by expanding in the
meson field momentum $P$. To order $P^2$, one obtains
\be
K_{11}^a(P) &=& Z_1^a (P^2 - M_1^{a 2} ),
\hspace{2em} K_{22}^a(P) \;\; = \;\; Z_2^a (P^2 - M_2^{a 2} )
\nonumber \\
K_{12}^a(P) &=& K_{21}^a(P) \;\; = \;\;
\gamma^a (P^2 - \Delta^2 \delta_{ab}|_{b = 4,...,7}),~~
(\Delta = m_s - m_u)
\label{K_matrix}
\ee
where
\be
Z_1^a &=& 4 I_2^a Z , \hspace{2em} Z_2^a \; = \; 4 I_2^{ff a} \bar{Z},
\hspace{2em} \gamma^a \; = \; 4 I_2^{f a} Z,
\label{I_12} \\
M_1^{a 2} &=& (Z_1^a)^{-1}[\frac{1}{G_1}-4(I_1^a(m_q^a) +
I_1^a(m_{q'}^a)] +
Z^{-1} \Delta^2 \delta_{ab}|_{b = 4,...,7}~ ,
\label{M_1} \\
M_2^{a 2} &=& (Z_2^a)^{-1}[\frac{1}{G_1}-4(I_1^{ff a}(m_q^a) +
I_1^{ff a}(m_{q'}^a)] + \bar{Z}^{-1} \Delta^2 \delta_{ab}|_{b = 4,...,7}~ .
\label{M_2}
\ee
Here, $Z = 1 - \frac{6 m^2_u}{M^2_{a_1}} \approx 0.7$,
$\bar{Z} = 1 - \Gamma^2~\frac{6 m^2_u}{M^2_{a_1}} \approx 1$ (see eq.
(\ref{Gamma})),
$Z$ is the additional renormalization of the ground pseudoscalar meson
states, taking into account the $\phi^a \rightarrow A^a$ transitions
(see \cite{volk_97,VEN}).
$I_n^a, I_n^{f a}$ and $I_n^{ff a}$ denote the usual loop
integrals arising in the momentum expansion of the NJL quark
determinant, but now with zero, one or two factors
$f_a ({\bf k})$, eqs.(\ref{ffq}), in the
numerator (see (\ref{I_1^ff}) and below )
\be
I_2^{f..f a} (m_q, m_{q'}) &=& -i N_{\rm c}
\kint \frac{f_a(\bfk )..f_a(\bfk )}{(m_q^{a 2} - k^2)
(m_{q'}^{a 2} - k^2)}.
\label{I_2^ff}
\ee

After the renormalization of the meson fields
\be
\phi_i^{a r} = \sqrt{Z_i^a} \phi_i^a
\label{phi^r}
\ee
the part of the Lagrangian
(\ref{L_2}), describing the pions and kaons, takes the form
\be
L_\pi^{(2)} &=& \frac{1}{2} \left[ (P^2 - M^2_{\pi_1})~ \pi^2_1 +
2 \Gamma_\pi P^2~ \pi_1 \pi_2 + (P^2 - M^2_{\pi_2})~ \pi^2_2 \right],
\label{Lp}\\
L_K^{(2)} &=& \frac{1}{2} [ (P^2 - M^2_{K_1} - \Delta^2)~ K^2_1
+ (P^2 - M^2_{K_2} - \Delta^2)~ K^2_2 \nonumber \\
&+& 2 \Gamma_K (P^2 - \Delta^2)~ K_1 K_2 ].
\label{LK}
\ee
Here
\be
\Gamma_a &=& \frac{\gamma_a}{\sqrt{Z_1^a Z_2^a}} =
\frac{I_2^{f a} \sqrt{Z}}{\sqrt{I_2^a I_2^{ff a} \bar{Z}}}.
\label{Gamma}
\ee
After the transformations of the meson fields
\be
\phi^a
&=& \cos( \theta_a - \theta_a^0) \phi_1^{ar}
- \cos( \theta_a + \theta_a^0) \phi_2^{ar},   \nonumber \\
\phi^{'a}
&=& \sin ( \theta_a - \theta_a^0) \phi_1^{ar}
- \sin ( \theta_a + \theta_a^0) \phi_2^{ar}
\label{transf}
\ee
the Lagrangians (\ref{Lp}) and (\ref{LK}) take the diagonal forms
\be
L_\pi^{(2)} &=& \half (P^2 - M_\pi^2)~ \pi^2 +
\half (P^2 - M_{\pi'}^2)~ \pi^{' 2}, \\
L_K^{(2)} &=& \half (P^2 - M_K^2)~ K^2 +
\half (P^2 - M_{K'}^2)~ K^{' 2}.
\label{L_pK}
\ee
Here
\be
&&M^2_{(\pi, \pi')} = \frac{1}{2 (1 - \Gamma^2_\pi)}
[M^2_{\pi_1} + M^2_{\pi_2} \nonumber \\
&&\qquad (- , +)~ \sqrt{(M^2_{\pi_1} - M^2_{\pi_2})^2 +
(2 M_{\pi_1} M_{\pi_2} \Gamma_\pi)^2}], \\
&&M^2_{(K, K')} = \frac{1}{2 (1 - \Gamma^2_K)} [M^2_{K_1} + M^2_{K_2}
+ 2 \Delta^2 (1 - \Gamma^2_K)  \nonumber \\
&&\qquad (- , +)~ \sqrt{(M^2_{K_1} - M^2_{K_2})^2 +
(2 M_{K_1} M_{K_2} \Gamma_K)^2}].
\label{MpK}
\ee
and
\be
\tan 2 \bar{\theta_a} = \sqrt{\frac{1}{\Gamma^2_a} -
1}~\left[ \frac{M^2_{\phi_1^a}
- M^2_{\phi_2^a}}{M^2_{\phi_1^a} + M^2_{\phi_2^a}} \right],~~~~~~
2 \theta_a = 2 \bar{\theta_a} + \pi.
\label{tan}
\ee
\be
\sin \theta_a^0 =\sqrt{{1+\Gamma_a}\over 2}
\label{theta0}
\ee
\be
M_{\pi_1}^2 &=& (4 Z I_2(m_u, m_u))^{-1}[\frac{1}{G_1}-8 I_1(m_u)] =
\frac{m_u^0}{4 Z m_u I_2 (m_u, m_u) G_1}, \nonumber\\
M_{\pi_2}^2 &=& (4 I_2^{ff}(m_u, m_u))^{-1}[\frac{1}{G_1}-
8 I_1^{ff}(m_u)],
\label{Mp}
\ee
\be
M_{K_1}^2 &=& (4 Z I_2(m_u, m_s))^{-1}[\frac{1}{G_1} -
4 (I_1(m_u) + I_1(m_s))] + (Z^{-1} - 1) \Delta^2  \nonumber \\
&=& \frac{\frac{m_u^0}{m_u} + \frac{m_s^0}{m_s}}{4 Z
I_2(m_u, m_s) G_1} + (Z^{-1} - 1) \Delta^2, \nonumber\\
M_{K_2}^2 &=& (4 I_2^{ff}(m_u, m_s))^{-1}[\frac{1}{G_1} -
4 (I_1^{ff}(m_u) + I_1^{ff}(m_s))].
\label{MK}
\ee

\section{The effective Lagrangian for the ground and excited
states of the vector mesons}
The free part of the effective Lagrangian (\ref{12}) describing
the ground and excited states of the vector mesons has the form
\be
L^{(2)} (V) &=&
- \half \sum_{i, j = 1}^{2} \sum_{a = 0}^{8} V_i^{\mu a} (P)
R_{ij}^{\mu\nu a} (P) V_j^{\nu a}(P) ,
\label{LV_2}
\ee
where
\be
\sum_{a = 0}^{3} V_i^{\mu a} = (\omega_i^\mu)^2 +
(\rho_i^{0 \mu})^2 +
2 \rho_i^{+ \mu} \rho_i^{- \mu},~~~
(V_i^{4 \mu})^2 + (V_i^{5 \mu})^2 = 2 K_i^{* + \mu} K_i^{* - \mu}, \nonumber \\
(V_i^{6 \mu})^2 + (V_i^{7 \mu})^2 = 2 K_i^{* 0 \mu}
K_i^{* 0 \mu},~~~ (V_i^{8 \mu})^2 = (\varphi_i^\mu)^2~~~~~~~~~~
\label{V^a}
\ee
and
\be
R_{ij}^{\mu \nu a} (P) =
-~ \frac{\delta_{ij}}{G_2} g^{\mu\nu}
-~ i~ N_{\rm c} \; {\rm tr}\, \kint \left[
\frac{1}{\kslash + \half\Pslash - m_q^a}\gamma^\mu f_i^{a,V}
\frac{1}{\kslash - \half\Pslash - m_{q'}^a}  \gamma^\nu f_j^{a,V}
\right]  , \nonumber \\
f_1^{a,V} \equiv 1, \hspace{2em} f_2^{a,V} \;\; \equiv \;\; f_a^V ({\bf k}).\hspace{3cm}
\label{R_full}
\ee
To order $P^2$, one obtains
\be
R_{11}^{\mu\nu a} &=& W_1^a [P^2 g^{\mu\nu} - P^\mu P^\nu -
g^{\mu\nu} (\bar M^a_1)^2], \nonumber \\
R_{22}^{\mu\nu a} &=& W_2^a [P^2 g^{\mu\nu} - P^\mu P^\nu -
g^{\mu\nu} (\bar M^a_2)^2], \nonumber \\
R_{12}^{\mu\nu a} &=& R_{21}^{\mu\nu a} = \bar\gamma^a
[P^2 g^{\mu\nu} - P^\mu P^\nu - \frac{3}{2} \Delta^2 g^{\mu\nu}
\delta^{ab}|_{b = 4..7}].
\label{R_ij}
\ee
Here
\be
W_1^a &=& \frac{8}{3} I_2^a,~~~W_2^a = \frac{8}{3} I_2^{ff a},~~~
\bar\gamma^a = \frac{8}{3} I_2^{f a}, \label{W} \\
(\bar M_1^a)^2 &=& (W_1^a G_2)^{-1} + \frac{3}{2}
\Delta^2 \delta^{ab}|_{b = 4..7}, \\
(\bar M_2^a)^2 &=& (W_2^a G_2)^{-1} + \frac{3}{2}
\Delta^2 \delta^{ab}|_{b = 4..7}.
\label{WM}
\ee
After renormalization of the meson fields
\be
V_i^{\mu a r} = \sqrt{W_i^a}~V_i^{\mu a}
\label{V^r}
\ee
we obtain the Lagrangians
\be
L_\rho^{(2)} &=& - \half [( g^{\mu\nu} P^2 - P^\mu P^\nu -
g^{\mu\nu} M^2_{\rho_1}) \rho^\mu_1 \rho^\nu_1 \nonumber \\
&+& 2 \Gamma_\rho  ( g^{\mu\nu} P^2 - P^\mu P^\nu) \rho_1^\mu
\rho_2^\nu + ( g^{\mu\nu} P^2 - P^\mu P^\nu -
g^{\mu\nu} M^2_{\rho_2}) \rho^\mu_2 \rho^\nu_2 ],
\label{L2_V1}
\ee
\be
L_\varphi^{(2)} &=& - \half [( g^{\mu\nu} P^2 - P^\mu P^\nu -
g^{\mu\nu} M^2_{\phi_1}) \varphi^\mu_1 \varphi^\nu_1 \nonumber \\
&+& 2 \Gamma_\phi  ( g^{\mu\nu} P^2 - P^\mu P^\nu) \varphi_1^\mu
\varphi_2^\nu + ( g^{\mu\nu} P^2 - P^\mu P^\nu -
g^{\mu\nu} M^2_{\varphi_2}) \varphi^\mu_2 \varphi^\nu_2 ],
\label{L2_V2}
\ee
\be
L_{K^*}^{(2)} &=& - \half [( g^{\mu\nu} P^2 - P^\mu P^\nu -
g^{\mu\nu} (\frac{3}{2} \Delta^2 + M^2_{K^*_1})) K^{*\mu}_1
K^{*\nu}_1 \nonumber \\
&+& 2 \Gamma_{K^*} ( g^{\mu\nu} P^2 - P^\mu P^\nu -
g^{\mu\nu} \frac{3}{2}
\Delta^2) K_1^{*\mu} K_2^{*\nu} \nonumber \\
&+& ( g^{\mu\nu} P^2 - P^\mu P^\nu -
g^{\mu\nu} (\frac{3}{2} \Delta^2 + M^2_{K^*_2})) K^{*\mu}_2 K^{*\nu}_2 ].
\label{L2_V3}
\ee
Here
\be
M_{\rho_1}^2 = \frac{3}{8 G_2 I_2(m_u, m_u)},~~~
M_{{K^*}_1}^2 = \frac{3}{8 G_2 I_2(m_u, m_s)},  \nonumber \\
M_{\phi_1}^2 = \frac{3}{8 G_2 I_2(m_s, m_s)},~~~
M_{\rho_2}^2 = \frac{3}{8 G_2 I^{ff}_2(m_u, m_u)}, \nonumber \\
M_{{K^*}_2}^2 = \frac{3}{8 G_2 I^{ff}_2(m_u, m_s)},~~~
M_{\phi_2}^2 = \frac{3}{8 G_2 I^{ff}_2(m_s, m_s)},
\label{MV_i}
\ee
\be
\Gamma_{a_{i,j}} = \frac{I_2^{f a}(m_i, m_j)}
{\sqrt{I_2^a(m_i, m_j)I_2^{ff a}(m_i, m_j)}}.
\label{GammaV}
\ee
After transformations of the vector meson fields, similar to
eqs.~(\ref{transf}) for the pseudoscalar mesons, the Lagrangians
(\ref{L2_V1},\ref{L2_V2},\ref{L2_V3}) take the diagonal form
\be
L^{(2)}_{V^a, \bar V^a} = - \half \left[ (g^{\mu\nu} P^2 -
P^\mu P^\nu - M^2_{V^a} ) V^{a \mu} V^{a \nu}  \right. \nonumber \\
\left. + (g^{\mu\nu} P^2 - P^\mu P^\nu -
M^2_{\bar V^a} ) \bar V^{a \mu} \bar V^{a \nu} \right],
\label{LDV}
\ee
where $V^a$ and $\bar V^a$ are the physical ground and excited
states vector mesons
\be
\kern-2mm M^2_{\rho, \bar\rho} &=& \frac{1}{2 (1 - \Gamma^2_\rho)}
\left[M^2_{\rho_1} + M^2_{\rho_2} (- , +) \sqrt{(M^2_{\rho_1} -
M^2_{\rho_2})^2 + (2 M_{\rho_1}M_{\rho_2} \Gamma_\rho)^2}
\right],
\label{Mrho}
\ee
\be
M^2_{\phi, \bar\phi} = \frac{1}{2 (1 - \Gamma^2_\phi)} \left[
M^2_{\phi_1} + M^2_{\phi_2}~ (- , +)~ \sqrt{(M^2_{\phi_1} -
M^2_{\phi_2})^2 + (2 M_{\phi_1}M_{\phi_2}
\Gamma_\phi)^2} \right] ,
\label{Mphi}
\ee
\be
M^2_{K^*, \bar K^*} = \frac{1}{2 (1 - \Gamma^2_{K^*})} \left[
M^2_{K^*_1} + M^2_{K^*_2} + 3 \Delta^2 (1 - \Gamma^2_{K^*})
\right.\nonumber \\
\left.
~( - , + )~ \sqrt{(M^2_{K^*_1} - M^2_{K^*_2})^2 +
(2 M_{K^*_1}M_{K^*_2} \Gamma_{K^*})^2} \right].
\label{MK^*}
\ee

\section{Model parameters and meson masses}

In the paper \cite{volk_97}, there were obtained  numerical
estimations for the model parameters, meson masses and
weak decay constants in our model.
Here we only give the values:
the constituent quark
masses are $m_u\approx m_d\approx 280$ MeV,
$m_s\approx 455 $ MeV; the cut-off parameter $\Lambda_3=1.03$ GeV;
the four-quark coupling constants $G_1=3.47$ GeV$^{-2}$ and
$G_2=12.5$ GeV$^{-2}$;
the slope parameters in the form factors $d_{uu}=-1.784$ GeV$^{-2}$,
$d_{us}=-1.757$ GeV$^{-2}$, $d_{ss}=-1.727$ GeV$^{-2}$;
the external parameters in the form factors
$c_{uu}^{\pi}=1.37$, $c_{uu}^{\rho}=1.32$, $c_{us}^{K}=1.45$,
$c_{us}^{K^*}=1.54$ and $c_{ss}^{\varphi}=1.41$.

With the model parameters fixed, we  obtain the angles $\theta_a$
and $\theta_a^0$:
\be
\ba{lll}
\theta_{\pi}=59.48^{\circ} &
\theta_{\pi}^0=59.12^{\circ}, &
\theta_{\rho}=81.8^{\circ},\\
\theta_{\rho}^0=81.5^{\circ},&
\theta_{K}=60.2^{\circ},&
\theta_{K}^0=57.13^{\circ}\\
\theta_{K^{*}}=84.7^{\circ},&
\theta_{K^{*}}^0=59.14^{\circ},&
\theta_{\varphi}=68.4^{\circ},\\
\theta_{\varphi}^{0}=57.13^{\circ}
\ea
\ee
and the meson masses are
\be
\ba{lll}
M_{\pi}=136 $ MeV$, &
M_{\pi'}=1.3 $ GeV$, &
M_{\rho}=768$ MeV$, \\
M_{\rho'}=1.49 $ GeV$, &
M_{K}=496 $ MeV$, &
M_{K'}=1.45 $ GeV$, \\
M_{K^*}=887 $ MeV$,  &
M_{K^{*'}}=1.479 $ GeV$. &
\ea
\ee
The experimental values are  \cite{Rev_98}
\be
\ba{lll}
M_{\pi^{\pm}}^{exp}=134.9764\pm0.0006 $ MeV$, &
M_{\pi'}^{exp}=1300\pm 100 $ MeV$, &
M_{\rho}^{exp}=768.5\pm 0.6 $ MeV$, \\
M_{\rho'}^{exp}=1465\pm 25 $ MeV$, &
M_{K^{\pm}}^{exp}=493.677\pm 0.016 $ MeV$, &
M_{K'}^{exp}\approx 1460 $ MeV$, \\
M_{K^{*\pm}}^{exp}=891.59\pm 0.24 $ MeV$,  &
M_{K^{*'}}^{exp}=1412\pm 12 $ MeV$. &
\ea
\ee
For the weak decay constants we have
\be
\ba{ll}
F_{\pi}=93 $ MeV$, &
F_{\pi'}=0.57 $ MeV$,\\
F_{K}=108 $ MeV $\approx 1.16 F_{\pi}, &
F_{K'}=3.3$ MeV$.
\ea
\ee
Now we can calculate the strong decay widths of  $K'$ and $K^{*'}$.

\section{The decays $K^{*'}\to K\rho$, $K^{*'}\to K^*\pi$,
$K^{*'}\to K\pi$, $\varphi'\to K^* K$, $\varphi' \to \bar KK$}

\begin{figure}
\psfig{file=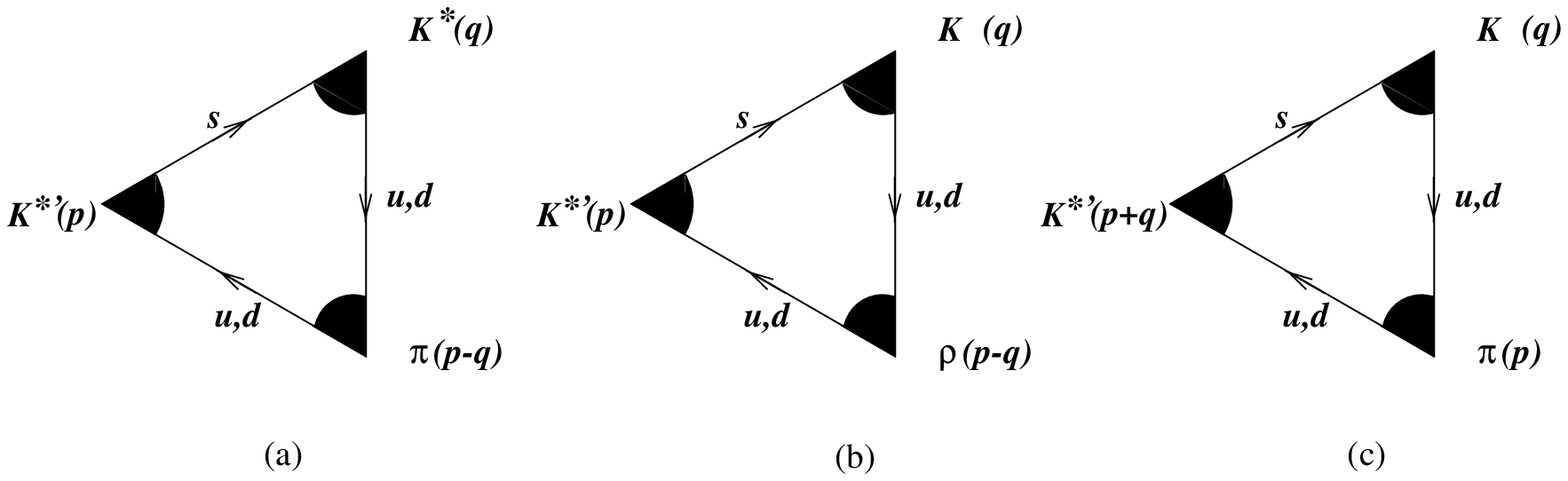}
\caption{The one-loop diagrams set for the decays of $K^{*\prime}$. }
\label{kstar}
\end{figure}

\begin{figure}
\psfig{file=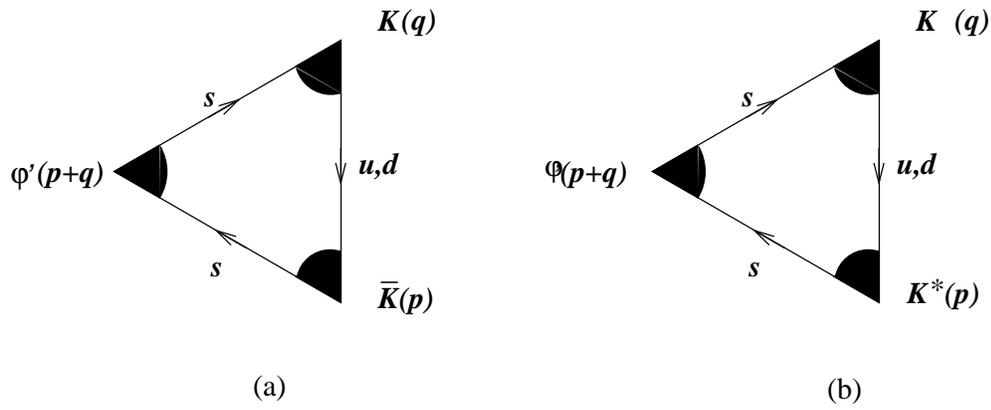}
\caption{The one-loop diagrams set for the decays of $\varphi'$. }
\label{phi}
\end{figure}

In the framework of our model, the  decay modes
of the excited strange vector mesons $K^{*'}$ and $\varphi'$
are represented by the triangle diagrams shown in
Figs.\ref{kstar} and \ref{phi}.
When calculating these diagrams, we keep the least possible
dependence on the external momenta: squared for the anomaly type
graphs and linear for  another type. We omit here the higher order
momentum dependence.

As it has been mentioned in this paper,
every vertex  is now momentum dependent and includes
form factors defined  in Sec.~3
(see eq.(\ref{ffq})).  In Figs.\ref{kstar} and \ref{phi}
the presence of form factors is marked by black shaded
angles in vertices.
  Each black shaded vertex with a pseudoscalar meson
is implied to contain
the following linear combination for the ground state:
	\be
	\bar f_a={1\over \sin 2\theta_{a}^{0}} \left[
	{\sin(\theta_a+\theta_{a}^{0}) \over \sqrt{Z_1^a}}+
	{\sin(\theta_a-\theta_{a}^{0}) \over \sqrt{Z_2^a}}f_a
	\right], \label{physformfactors1}
	\ee
and for an excited state ---
	\be
	\bar f_{a}^{\prime}={-1\over \sin 2\theta_{a}^{0}} \left[
	{\cos(\theta_a+\theta_{a}^{0}) \over \sqrt{Z_1^a}}+
	{\cos(\theta_a-\theta_{a}^{0}) \over \sqrt{Z_2^a}}f_a
	\right], \label{physformfactors2}
	\ee
where
$\theta_a$ and $\theta_{a}^{0}$ are the angles defined in Sec.~3
(see eqs.~(\ref{tan}) and (\ref{theta0})) and
$f_a$ is one of the form factors
defined in Sec.~3 (eq. (\ref{ffq})).
In the case of vector meson vertices, we have the same linear combinations
except that $Z_i^a$ are to be replaced by $W_i^a$ (\ref{W}), and
the related angles and form factor parameters must be chosen.

Now we can calculate the decay widths of the
excited mesons. Let us start with the process
$K^{*'}\to K^*\pi$.
The corresponding amplitude, $T_{K^{*'}\to K^{*}\pi}$, has the form
	\be
	T_{K^{*'}\to K^{*}\pi}&=& g_{K^{*'}\to K^*\pi}
	\epsilon_{\mu\nu\alpha\beta}p^{\alpha}q^{\beta}
	\xi^{\mu}(p|\lambda)\xi^{\nu}(q|\lambda'),
	\label{T_Kstar_to_Kstarpi}
	\ee
where $p$  and $q$ are the momenta of the $K^{*'}$- and $K^{*}$-mesons,
respectively,
and $g_{K^{*'}\to K^*\pi}$  is the (dimensional) coupling
constant, which follows from the combination of one-loop integrals
	\be
	g_{K^{*'}\to K^*\pi}=
	{8 m_s\over m_u^2-m_s^2}
	\left(
	I_2^{\bar f_{K^{*}}^\prime\bar f_{K^*}\bar f_{\pi}}(m_u)-
	I_2^{\bar f_{K^{*}}^\prime\bar f_{K^*}\bar f_{\pi}}(m_u,m_s)
	\right).
	\label{KstartoKstarpi}
	\ee
Note that in (\ref{KstartoKstarpi}) the integrals
 $I_2^{f\dots f}$ are defined in the same way as in Sec.~3, eq.(\ref{I_2^ff}),
except that the form factors $f$ in (\ref{I_2^ff}) are replaced by
the expressions of type
 (\ref{physformfactors1}) and (\ref{physformfactors2}).
$\xi_{\mu}(p|\lambda)$ and $\xi_{\nu}(q|\lambda')$
are polarized vector wave functions ($\lambda,\lambda'=1,2,3$).

Using (\ref{physformfactors1}) and (\ref{physformfactors2})
we expand the above expression and rewrite it in  terms of
$I_2^{ff}$ defined in~(\ref{I_2^ff}).  The result
is too lengthy, so we omit it here.
For the parameters given in Sec.~5 we find
	\be
	g_{K^{*'}\to K^*\pi}=3.6\, {\rm GeV}^{-1}
	\ee
and the decay width
	\be
	\Gamma_{K^{*'}\to K^*\pi}=
	{g_{K^{*'}\to K^*\pi}^2\over 4\pi}|\vec {\bf p}|^3=70\,
	 {\rm MeV}
	\ee
Here $|\vec {\bf p}|$ is the 3-momentum of a produced particle in
the rest frame of the decaying meson.
The lower limit for this value coming from
experiment is $\sim 91\pm 9$ MeV~\cite{Rev_98}.

A similar calculation has to be performed for the rest of the $K^{*'}$ decay
modes under consideration.
The coupling constant $g_{K^{*'}\to K\rho}$ is derived in the
same way as in (\ref{KstartoKstarpi}), with the only
difference that $\bar f_{\pi}$ and $\bar f_{K^*}$ are to be replaced
by $\bar f_{\rho}$ and $\bar f_{K}$.
The corresponding amplitude,
$T_{K^{*'}\to K\rho}$, takes the form
	\be
	T_{K^{*'}\to K\rho}&=& g_{K^{*'}\to K\rho}
	\epsilon_{\mu\nu\alpha\beta}p^{\alpha}q^{\beta}
	\xi^{\mu}(p|\lambda)\xi^{\nu}(p-q|\lambda'),
	\label{T_Kstar_to_Krho}
	\ee
where $p$ and $q$ are the momenta of $K^{*'}$- and $K$-mesons,
respectively, and
	\be
	g_{K^{*'}\to K\rho}&=& {8m_s\over m_u^2-m_s^2}
	\left(
	I_2^{\bar f_{K^{*}}^\prime\bar f_K\bar f_{\rho}}(m_u)-
	I_2^{\bar f_{K^{*}}^\prime\bar f_K\bar f_{\rho}}(m_u,m_s)
	\right).  \label{g_Kstar_to_Krho}
	\label{KstartoKrho}
	\ee
The corresponding decay width follows from (\ref{T_Kstar_to_Krho})
and  (\ref{KstartoKrho})
via integration of the squared module of the decay amplitude over
the phase space of the final state
	\be
	\Gamma_{K^{*'}\to K\rho}
		&=& {g_{K^{*'}\to K\rho}^2\over 4\pi} |\vec {\bf p}|^3.
	\ee
For the parameters given in Sec.~5 one has
\be
 && g_{K^{*'}\to K\rho}=3.2\, {\rm GeV}^{-1},
\ee
\be
&&\Gamma_{K^{*'}\to K\rho} =23 \,{\rm MeV}.
\ee
From  experiment,  the upper limit for this process is
$\Gamma_{K^{*'}\to K\rho}^{exp} < 16\pm 1.5$ MeV.

The process $K^{*'}\to K\pi$ is described by the
amplitude
	\be
	T_{K^{*'}\to K\pi}&=&i \frac{g_{K^{*'}\to K\pi}}{2}
	(p-q)^{\mu}\xi_{\mu}(p+q|\lambda),
	\ee
where $p$ and $q$ are the momenta of $\pi$ and $K$,
and $\xi_{\mu}$ is
the $K^{*'}$ wave function.
The coupling constant $g_{K^{*'}\to K\pi}$ is obtained by calculating
the one-loop integral
	\be
	g_{K^{*'}\to K\pi}=
	4 I_2^{\bar f_{K^{*}}^\prime\bar f_{K}\bar f_{\pi}}(m_u,m_s)
	= 2.3
	\label{KstartoKpi}
	\ee
and the decay width is
	\be
	\Gamma_{K^{*'}\to K\pi}
	&=&{g_{K^{*'}\to K\pi}^2|\vec {\bf p}|^3\over 8\pi M^2_{K^{*'}}}=
	24\, {\rm MeV}.
	\ee
The experimental value is
$15\pm 5$ MeV~\cite{Rev_98}.

The mesons with hidden strangeness ($\varphi'$) are treated in
the same way as $K^{*'}$.
We consider the two decay modes: $\varphi'\to KK^*$ and
$\varphi'\to \bar KK$.
Their amplitudes are
	\be
	T_{\varphi'\to KK^*}&=& g_{\varphi'\to KK^*}
	\epsilon_{\mu\nu\alpha\beta}p^{\alpha}q^{\beta}
	\xi^{\mu}(p+q|\lambda)\xi^{\nu}(p|\lambda')\\
	T_{\varphi'\to \bar KK}&=& i g_{\varphi'\to \bar KK}
	(p-q)_{\mu}
	\xi^{\mu}(p+q|\lambda).
	\ee
Here $\xi_{\mu}$ and $\xi_{\nu}$ are the
wave functions of the $\varphi'$- and $K^*$-mesons,
$p$ and $q$  are the momenta of the $K$- and $K^*$-mesons.
The related coupling constants are
	\be
	g_{\varphi'\to KK^*}&=&
	{8m_u\over m_s^2-m_u^2}
	\left(
	 I_2^{\bar f_{\varphi}^\prime\bar f_{K^*}\bar f_{K}}(m_s)-
	 I_2^{\bar f_{\varphi}^\prime\bar f_{K^*}\bar f_{K}}(m_u,m_s)
	\right),\\
	g_{\varphi'\to \bar KK}&=&
	4 I_2^{\bar f_{\varphi}^\prime\bar f\bar f_{K}}(m_s).
	\ee
Thus, the decay widths are estimated as
	\be
	\Gamma_{\varphi'\to KK^*}&=&90 \, {\rm MeV}, \\
	\Gamma_{\varphi'\to \bar KK}&=&10 \, {\rm MeV}.
	\ee
Unfortunately,
there are no reliable data from experiment on the partial decay
widths for $\varphi'\to KK^*$ and $\varphi'\to \bar KK$
except the total width of $\varphi'$ being estimated
as $150\pm 50$ MeV~\cite{Rev_98}.
However, the dominance of the
process $\varphi'\to KK^*$ is observed,
which is in agreement with our result.

\section{The decays $K' \to K^*\pi$, $K' \to K\rho$, $K'\to K\pi\pi$}

\begin{figure}
\psfig{file=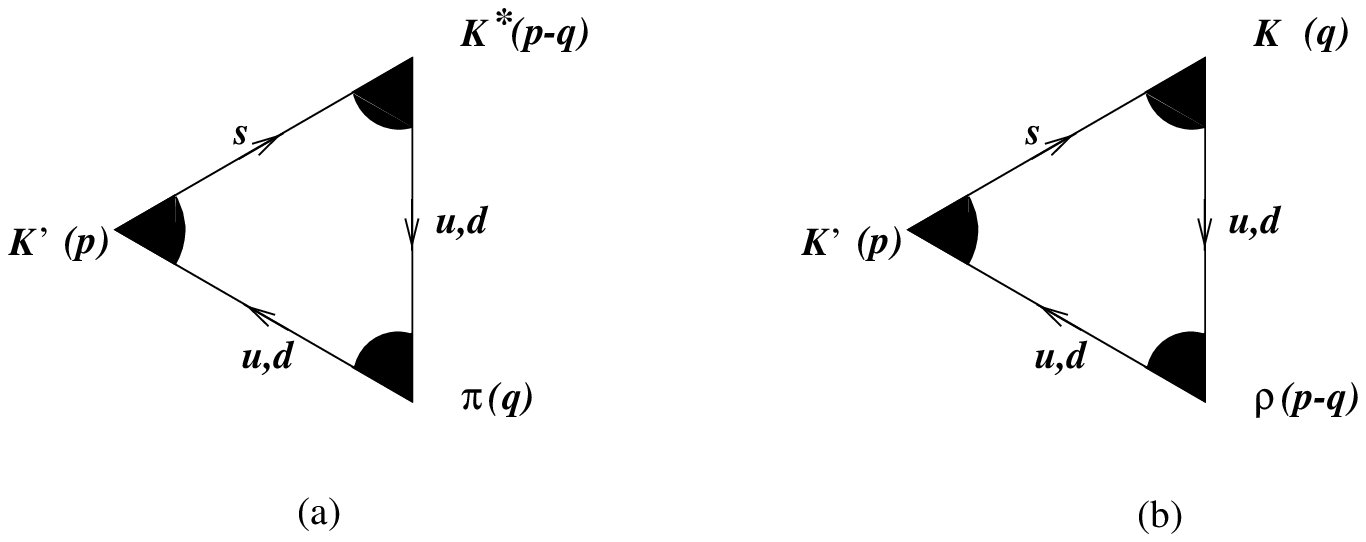}
\caption{The one-loop diagrams set for the decays of $K'$. }
\label{Kprime}
\end{figure}

Following the scheme outlined in the previous section,
we first estimate the $K'\to K^*\pi$ and $K'\to K\rho$ decay widths
(see Fig.\ref{Kprime}). Their amplitudes are
	\be
	T_{K'\to K^*\pi}&=&
	i g_{K'\to K^*\pi}(p+q)^{\mu}\xi_{\mu}(p-q|\lambda)\\
	T_{K'\to K\rho}&=&
	i g_{K'\to K\rho}(p+q)^{\mu}\xi_{\mu}(p-q|\lambda),
	\ee
here $p$ is the momentum of $K'$, $q$ is the momentum of $\pi$ ($K$),
and $\xi_{\mu}$ is
the vector meson wave function.
 The coupling constants are
	\be
	g_{K'\to K^*\pi}&=&
	4I_2^{\bar f_{K}^\prime\bar f_{K^*}\bar f_{\pi}}(m_u,m_s),\\
	g_{K'\to K\rho}&=&
	4 I_2^{\bar f_{K}^\prime\bar f_{K}\bar f_{\rho}}(m_u,m_s).
	\ee
By calculating the integrals in the  above
formulae we have $g_{K'\to K^*\pi}=-1.3$  and
$g_{K'\to K\rho}=-1.18$.
The decay widths thereby are
	\be
	\Gamma_{K'\to K^*\pi}&=&90\, {\rm MeV},\\
	\Gamma_{K'\to K\rho}&=&50\, {\rm MeV}.
	\ee
These processes have been seen in experiment and
the decay widths are~\cite{Rev_98}
	\be
	\Gamma^{exp}_{K'\to K^*\pi}&\sim &109\, {\rm MeV},\\
	\Gamma^{exp}_{K'\to K\rho}&\sim &34\, {\rm MeV}.
	\ee

\begin{figure}[t]
\psfig{file=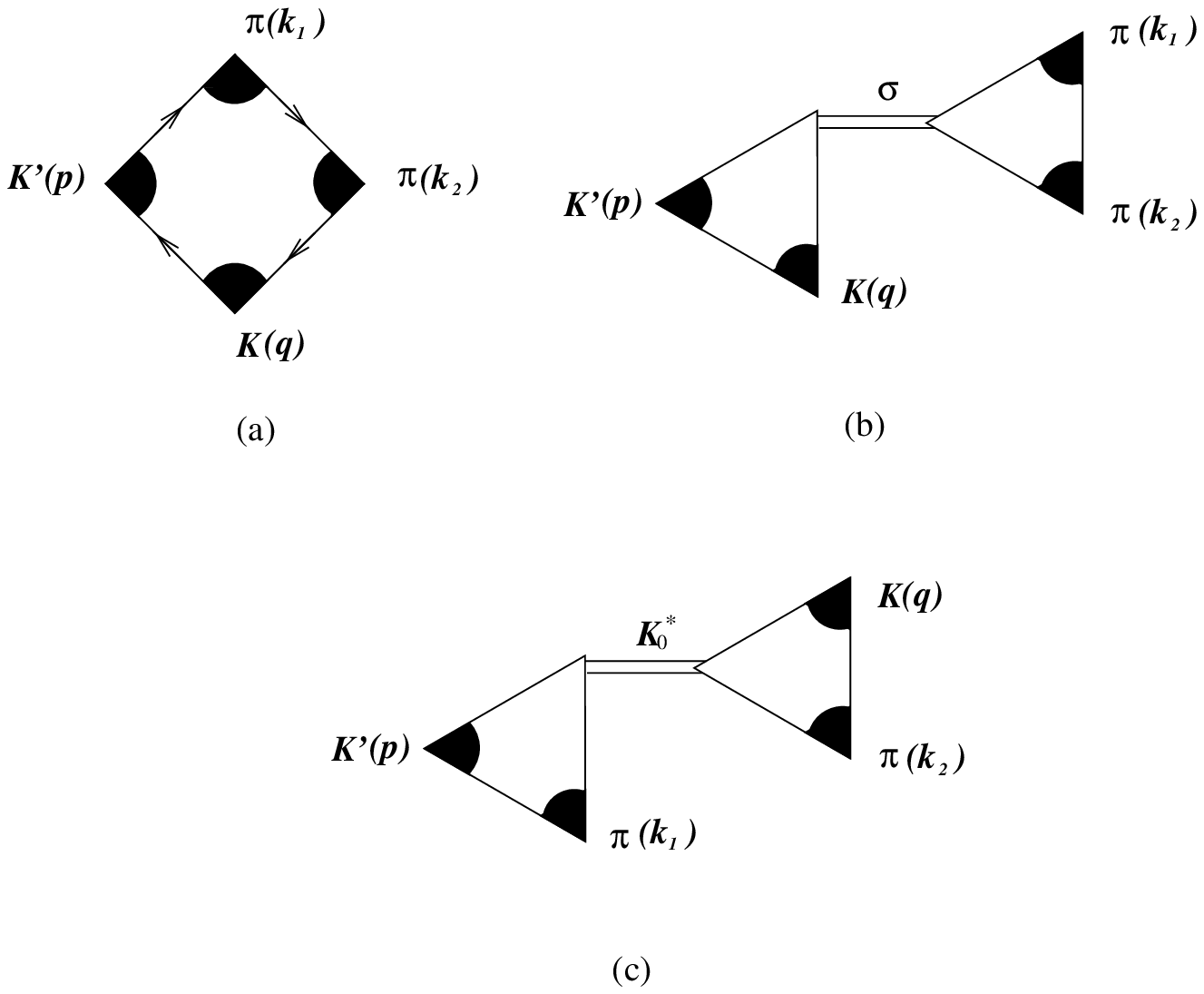}
\caption{The one-loop diagrams set for the decay  $K' \to K 2\pi$. }
\label{KtoK2pi}
\end{figure}

The remaining  decay $K'\to K\pi\pi$  into three particles
requires more complicated calculations.
In this case, one must consider a box diagram Fig.\ref{KtoK2pi}.(a)
and two types of diagrams Fig.\ref{KtoK2pi}.(b,c) with
intermediate $\sigma-$ and $K^*_0-$resonances.
The diagrams for resonant channels are approximated with
the use of the relativistic Breit-Wigner function.
The integration over the kinematically relevant range in
the phase space for final states gives
	\be
	\Gamma_{K'\to K\pi\pi}\sim 1 {\rm MeV}.
	\ee

\section{Summary and conclusions}
In the standard NJL model for the description of the interaction
of mesons, it is conventional to use the
one-loop quark approximation, where the external momentum
dependence in quark loops is neglected.
This allows one to obtain, in this approximation,
the chiral symmetric phenomenological Lagrangian
\cite{volkov_83,volk_86, ebert_86,voglweise,klev},
which gives a good description
 of the low-energy meson physics
in the energy range of an order of 1 GeV \cite{gasiorowicz}.
In this paper, we have used a similar method for  describing
the interaction of the excited mesons.
Insofar as the masses of the excited mesons noticeably exceed 1 GeV,
we pretend only to a qualitative description rather
than  quantitative agreement with the experiment.
For the light excited mesons, we have got the results  closer
to the experiment \cite{VEN}, while for  heavier
strange mesons we are only in qualitative agreement with
experimental data. One should note that
the description  of  all the decays has been obtained  without
introducing new parameters, besides those used for the
description of the mass spectrum.

In this work, we have shown that the dominant decays of the
excited vector mesons $K^{*'}$ and $\varphi'$
are the decays $K^{*'}\to K^*\pi (\rho K)$ and $\varphi'\to KK^*$,
which go through the triangle quark loops of the anomaly type.
These results are close to the experimental
ones \cite{Rev_98}. The decays of the type $K^{*'}\to K\pi$ and
$\varphi'\to K\bar K$, going through the other (not anomaly type) quark
diagrams,  have smaller decay widths, which is also in agreement
 with  experiment.

On the other hand, the main decays of the $K'$ meson $K'\to K^*\pi$,
$K'\to K\rho$ and $K'\to K\pi\pi$ are described
by the quark diagrams, which are similar to those for the decay
$\pi'\to\rho\pi$ (see \cite{VEN}).
The dominant decays here are the decays $K'\to K^*\pi$
and $K'\to K\rho$. These results do not contradict (qualitatively)
to the recent experimental data.
So one can see that our model  satisfactorily describes
not only the masses and weak decay constants of the
radially excited mesons \cite{volk_96, volk_97}
but also their decay widths.

We would like to emphasize once more that we did not use any additional
parameter for description of the decays~\cite{VEN}.
The model is too simple to pretend to a more exact
quantitative description of the meson decay widths.

A similar calculation has also been made in the $3P_1$ potential
model \cite{geras}.
Nonlocal versions of
chiral quark models for the description of
excited mesons states have been also  considered in various
works (see, for instance, \cite{AAA,Nedelko}).
In \cite{Shakin}
a generalized NJL model including a relativistic model of
confinement was used to study
the radial excitations of pseudoscalar and vector mesons.

In our further work we are going to describe the masses
and decay widths of the excited  states of $\eta$ and $\eta'$
mesons.

\section*{\centerline{Acknowledgment}}

The work is supported by the Heisenberg--Landau Program 1998
and by Grant RFFI No.~98-02-16135.

\newpage
\appendix

\end{document}